\documentclass[reprint,superscriptaddress,amsmath,amssymb,aps]{revtex4-2}

\usepackage{graphicx}
\usepackage{dcolumn}
\usepackage{bm}
\usepackage{physics}
\usepackage{gensymb}
\usepackage{hyperref}
\hypersetup{      
    urlcolor=cyan,
}

\begin{document}

\preprint{APS/123-QED}

\title{Jones matrix holography with metasurfaces}

\author{Noah A. Rubin}
\affiliation{John A. Paulson School of Engineering and Applied Sciences, Harvard University, Cambridge, MA 02138, USA}
\author{Aun Zaidi}
\affiliation{John A. Paulson School of Engineering and Applied Sciences, Harvard University, Cambridge, MA 02138, USA}
\author{Ahmed Dorrah}
\affiliation{John A. Paulson School of Engineering and Applied Sciences, Harvard University, Cambridge, MA 02138, USA}
\author{Zhujun Shi}
\affiliation{Department of Physics, Harvard University, Cambridge, MA 02138, USA}
\affiliation{Apple Inc. Cupertino, CA 95014, USA}
\author{Federico Capasso}
\email{capasso@seas.harvard.edu}
\affiliation{John A. Paulson School of Engineering and Applied Sciences, Harvard University, Cambridge, MA 02138, USA}

\date{\today}

\begin{abstract}
We propose a new class of computer generated holograms whose far fields possess designer-specified polarization response. We dub these Jones matrix holograms. We provide a simple procedure for their implementation using form-birefringent metasurfaces. Jones matrix holography generalizes a wide body of past work with a consistent mathematical framework, particularly in the field of metasurfaces, and suggests previously unrealized devices, examples of which are demonstrated here. In particular, we demonstrate holograms whose far-fields implement parallel polarization analysis and custom waveplate-like behavior.
\end{abstract}

\maketitle

\begin{figure}[t]
    \centering
    \includegraphics[width=\columnwidth]{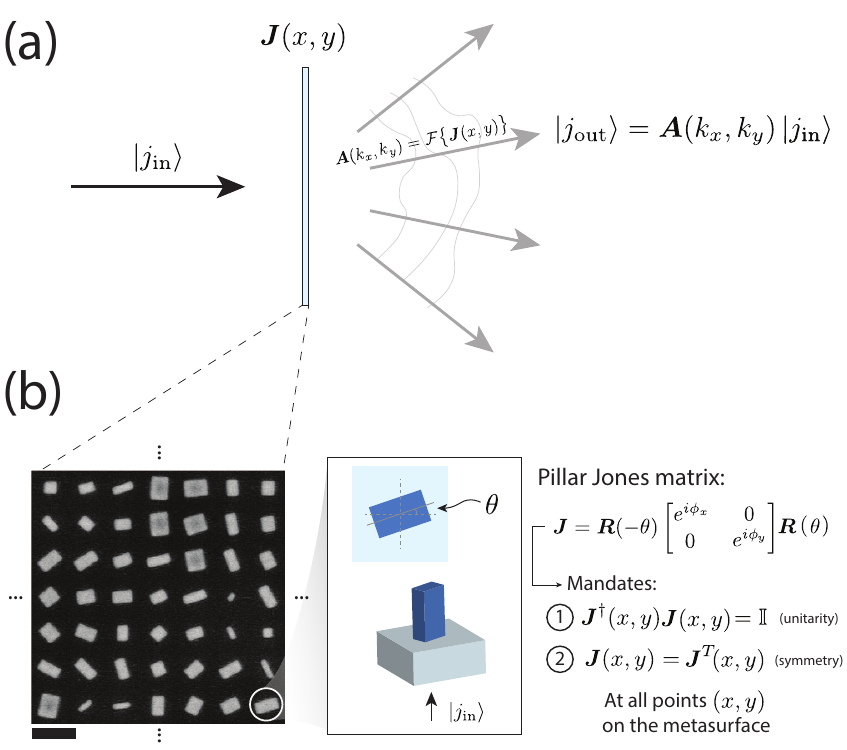}
    \caption{\textbf{(a)} A Jones matrix hologram implements a polarization-dependent mask ($\boldsymbol{J}(x,y)$) with a far-field, plane wave spectrum polarization response ($\boldsymbol{A}(k_x,k_y)$) whose behavior can be controlled. The far-field response to an incident polarization state with Jones vector $\ket{j_{\text{in}}}$ can be found by matrix multiplication. \textbf{(b)} These can be implemented with metasurfaces comprised of dielectric pillars, whose Jones matrices, as discussed in the text, are subject to certain mathematical constraints. An example SEM (scanning electron micrograph) of a section of a sample from this work is shown (scale bar 500 nm).}
    \label{fig:fig1}
\end{figure}

\section{Introduction}

In a 1965 paper~\cite{Lohmann1965}, Adolf Lohmann, a pioneer of the computer generated hologram (CGH), remarked that while ``holography'' translates as ``total recording'' from Greek, ``a hologram is not really a total recording, since only one amplitude and one phase are recorded, which would be adequate if light were a scalar wave.'' Indeed, light's polarization is often omitted in the study of holography and diffractive optics. However, a variety of holographic materials and technologies do permit the control of polarization in a spatially-varying fashion. Over the past five or-so decades, these have taken on various forms and names, among them polarization holograms~\cite{Todorov1984, Nikolova2009}, polarization gratings~\cite{Cincotti2003, Hasman2005}, a variety of liquid crystal devices, and, more recently, metasurfaces~\cite{Arbabi2015, Mueller2017}. These metasurfaces, subwavelength spaced-arrays of phase shifting elements which may be strongly form-birefringent, are the specific focus and implementation medium of this work. However, the generalized viewpoint we discuss here has broader applicability.

In the paraxial regime---often an unstated assumption in holography and Fourier optics---a propagator, commonly the Fourier transform $\mathcal{F}$, links the ``near-field''---an electric field with a phase and/or amplitude distribution created by the hologram (though not to be confused with the optical near field owed to evanescent waves)---with the ''far-field'', a desired phase and/or amplitude (though more commonly, amplitude) distribution some distance many wavelengths away. In this work, for clarity's sake, ``hologram'' refers to the physical field-modifying object, rather than a holographic image in the far-field which might also be more colloquially referred to as a ``hologram''. Often, a hologram is described by its spatially-varying, complex-valued aperture transmission (or reflection) function $t(x,y)$, a single complex scalar function given by an amplitude and a phase. For normally incident, plane-wave-like light, $t(x,y)$ can be used as a stand-in for the field itself (an assumption that, if necessary, is easily relaxed with the convolution theorem). This picture can be generalized to handle polarization by describing the hologram instead by a $2\times2$ Jones matrix transfer function $\boldsymbol{J}(x,y)$ (matrix-valued quantities are denoted by bold lettering here), permitting the analysis of polarization-sensitive holographic media like those above. $\boldsymbol{J}(x,y)$, which describes the polarization response at each point $(x,y)$, contains four complex numbers, in contrast to the single complex number $t(x,y)$.

In a common approach, rather than using this Jones matrix description, the response of a metasurface (or other polarization-sensitive holographic element) is considered separately upon illumination with one of two orthogonal ``basis'' polarization states, which can be elliptical in general. An incident plane wave in one of the basis states, after passing through the metasurface, is designed to create a scalar field that is everywhere uniform in polarization, with a designer-specified overall phase profile. This approach permits the realization of optical elements---gratings, lenses, and in general, holograms---whose far-field function can switch on the basis of incident polarization (as in~\cite{Arbabi2015, Mueller2017}, to name only a few examples from what is now a vast literature). However, this switchability is global in nature: One entire far-field response is ascribed to each polarization state in the chosen basis, and the polarization-dependent response cannot vary over the far field. For all other polarizations, the response is a weighted superposition of the two. In other words, what is inherently a polarization-dependent problem is in this scheme reduced to two scalar ones (imparting two scalar phase profiles) --- accordingly, we dub this the ``scalar'' approach.

In another now-common approach, the metasurface is designed to produce a distribution of polarization ellipses, described by the Jones vector function $\ket{j(x,y)}$, so that the far field is given by $\ket{a(k_x, k_y)}=\mathcal{F}\{\ket{j(x,y)}\}$ where $\mathcal{F}$ now denotes the Fourier transform operator distributed over both elements of the Jones vector (Jones vectors are denoted throughout by kets in the bra-ket notation). In this way, the polarization state of the far-field can be made to vary in a desired fashion \cite{Deng2018, Rubin2018, Arbabi2019, Zhao2018, Ding2020}. However, this \emph{vectorial} approach assumes that the incident light has a particular polarization state. If this changes, so too does the carefully choreographed far-field polarization distribution.

\section{Jones matrix holography}

\subsection{Concept}

Neither the scalar nor vector approaches recognize the most general polarization-control enabled by the ability to spatially manipulate light's polarization (e.g., with a metasurface). In this work, we rely instead on the top-level design of the metasurface as a spatially-varying Jones matrix, specified \emph{without} regard for any particular incident polarization state. The metasurface is then described by a spatially-varying Jones matrix $\boldsymbol{J}(x,y)$ and a far-field $\boldsymbol{A}(k_x, k_y)=\mathcal{F}\{\boldsymbol{J}(x,y)\}$ where the Fourier transform now distributes over all four elements (four complex-valued functions) of the Jones matrix. If we assume plane wave incidence, $\boldsymbol{A}(k_x, k_y)$---itself a Jones matrix---gives the polarization-dependent behavior of each plane wave $(k_x, k_y)$ component of the far-field (Fig. \ref{fig:fig1}(a)). In other words, in this approach, rather than trying to control the far-field's intensity for some incident polarization or its polarization state, we seek to control its polarization \emph{transfer-function}. For instance, if $\boldsymbol{A}(k_x, k_y)$ corresponds to an $x$ polarizer, light will be directed into the plane wave component with direction $(k_x, k_y)$ in a way that depends on the incident polarization state in accordance with a polarizer --- bright if it is $\ket{x}$, dark if it is $\ket{y}$.

Importantly, this matrix approach encompasses the aforementioned vector and scalar ones as subcases: $\ket{j(x,y)}=\boldsymbol{J}(x,y)\ket{\eta}$ is a distribution of polarization states for a chosen incident polarization $\ket{\eta}$, and $\braket{j(x,y)}{\kappa}$ is a scalar field for a chosen analysis polarization state $\ket{\kappa}$. This approach thus provides a simple language in which to frame this problem and classify past works.

Now, if a far-field with a polarization-dependent response described by some $\boldsymbol{A}(k_x, k_y)$ is desired, a \emph{Jones matrix hologram} implementing it is given by inverse Fourier transform as

\begin{equation}
\label{eq:inverse_ft}
    \boldsymbol{J}(x,y) = \mathcal{F}^{-1}\{\boldsymbol{A}(k_x, k_y)\}.
\end{equation}

The $\boldsymbol{J}$ so-obtained and, for that matter, any Jones matrix can be decomposed as

\begin{equation}
\label{eqn:matrix_polar}
    \boldsymbol{J} = \boldsymbol{H}\boldsymbol{U}
\end{equation}

 where $\boldsymbol{H}$ is a Hermitian (lossy, polarizer-like) Jones matrix with $\boldsymbol{H}^{\dagger} = \boldsymbol{H}$ and $\boldsymbol{U}$ is a unitary (lossless, waveplate-like) Jones matrix with $\boldsymbol{U}^{\dagger}\boldsymbol{U} = \mathbb{I}$ ($\mathbb{I}$ being the $2\times2$ identity matrix and $\dagger$ the Hermitian conjugate). This is known as the matrix polar decomposition, derived from the more common singular value decomposition, and is the matrix analogue of the scalar polar decomposition of a number into amplitude and phase with $\boldsymbol{U}$ playing the role of a phasor and $\boldsymbol{H}$ the role of an amplitude~\cite{Chipman2019}. In general, the ``near-field'' $\boldsymbol{J}(x,y)$ corresponding to a desired ``far-field'' $\boldsymbol{A}(k_x,k_y)$ by Eq. \ref{eq:inverse_ft} will be neither strictly Hermitian nor unitary.

\subsection{Metasurface implementation}

To extend beyond mathematics into the regime of application, however, the Jones matrix function $\boldsymbol{J}(x,y)$ must be physically realizable as an optic. In this work, this is accomplished with metasurfaces comprised of dielectric pillars. In a now common implementation, these pillars are everywhere uniform in height (for ease of fabrication), with cross-sections possessing two perpendicular mirror symmetry axes (e.g., rectangles or ellipses). These metasurfaces exhibit form-birefringence, implementing a local Jones matrix of the form

\begin{equation}
\label{eq:metasurface_jones matrix}
    \boldsymbol{J}(x,y) = \boldsymbol{R}(-\theta)\begin{bmatrix} e^{i\phi_{x^{\prime}}} & 0 \\ 0 & e^{i\phi_{y^{\prime}}} \end{bmatrix}\boldsymbol{R}(\theta).
\end{equation}

where $\phi_{x^{\prime}}$ and $\phi_{y^{\prime}}$ are phases imparted on light polarized along the symmetry axes of the pillar, controlled by varying the elements' transverse dimensions, and $\theta$ is its angular orientation~\cite{Arbabi2015} (Fig. \ref{fig:fig1}(b)). Notably, Eq. \ref{eq:metasurface_jones matrix} has just three scalar degrees-of-freedom, while an arbitrary Jones matrix (with its four complex entries) has eight. A pillar-based dielectric metasurface, then, cannot implement \emph{any} desired Jones matrix from point-to-point. Eq. \ref{eq:metasurface_jones matrix} instead describes a Jones matrix subject to two key restrictions, those being 1) unitarity, such that $\boldsymbol{J}^{\dagger}\boldsymbol{J}=\mathbb{I}$ and, 2) symmetry, such that $\boldsymbol{J}^T=\boldsymbol{J}$ where $T$ denotes a matrix transpose. The former is a matrix generalization of the statement for scalar light that certain holographic media are phase- or amplitude-only. The latter is a way of stating that the eigen-polarizations of Eq. \ref{eq:metasurface_jones matrix} must be linear (no chirality).

These two restrictions imposed by this particular metasurface platform themselves impose restrictions on the far-field polarization function $\boldsymbol{A}(k_x, k_y)$ achievable with such a metasurface. Any desired far-field function $\boldsymbol{A}_{\text{des}}(k_x, k_y)$ we ask of the metasurface must first be compatible with these. The first restriction, matrix symmetry is easily accounted for: $\boldsymbol{J}(x,y)$ will only be symmetric if the far-field behavior specified by $\boldsymbol{A}(k_x, k_y)$ is everywhere a symmetric Jones matrix, too, since the matrix Fourier transform linking the two, being essentially an (infinitesimal, exponential-weighted) summation, preserves matrix symmetry. A metasurface whose local Jones matrix is of the form of Eq. \ref{eq:metasurface_jones matrix}, then, may only implement polarization behavior in the far-field described by symmetric Jones matrices; some practical consequences of this rule are elaborated below.

\begin{figure}[t]
    \centering
    \includegraphics[width=\columnwidth]{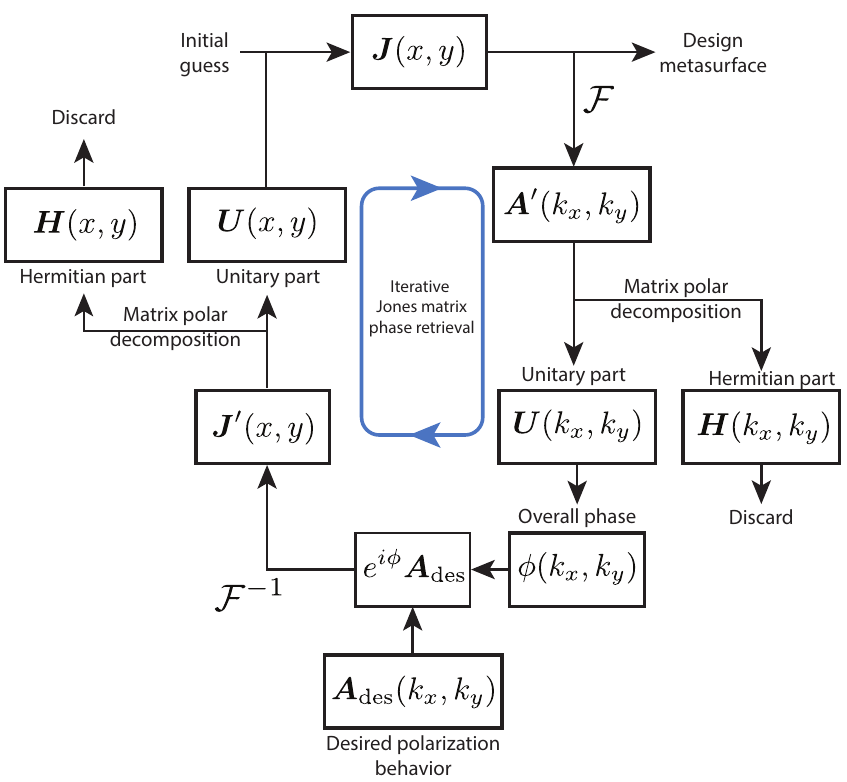}
    \caption{\textbf{Jones matrix phase retrieval:} The Gerchberg-Saxton (GS) algorithm can be generalized to allow a unitary Jones matrix mask to implement a far-field whose polarization-dependence $\boldsymbol{A}{(k_x, k_y)}$ may be arbitrary. Much like the scalar GS algorithm, this matrix generalization uses a division into phase-like (unitary) and amplitude-like (Hermitian) components using the matrix polar decomposition.}
    \label{fig:fig2}
\end{figure}

However, in general, even if we limit ourselves to a desired far-field behavior $\boldsymbol{A}(k_x, k_y)$ that is symmetric everywhere, the hologram $\boldsymbol{J}(x,y)$ obtained by Eq. \ref{eq:inverse_ft} will contain both Hermitian and unitary behavior in general, incompatible with the second restriction imposed by Eq. \ref{eq:metasurface_jones matrix}. This issue is less easily addressed. However, progress can be made by recognizing the problem's simpler, scalar analogue: the phase problem. 

\subsection{Jones matrix phase retrieval}

The phase problem describes the inverse problem of finding a scalar function that is strictly phase-only which may, nonetheless, have a Fourier transform (far-field) whose amplitude can be arbitrarily specified (while its phase is allowed to freely vary). This problem is well-known in holography, where the Fourier transform is over space, but also appears in time domain problems, for instance in determining modal phases from autocorrelation in lasers~\cite{Trebino2000}.

The phase problem is inherently one of variational calculus, of finding a functional $\phi(x,y)$ such that $\mathcal{F}\{e^{i\phi(x,y)}\}$ is optimum with respect to some function and given constraints. Rather than resorting to a brute-force gradient optimization (which would be numerically impractical over the scale of a large CGH, where the phase at each grid location becomes a free parameter), a number of ``gradient-free'' numerical techniques have emerged that are in wide use in diffractive optics (see, e.g.,~\cite{Arrizon2007, Rosales-Guzman2017}). By far the best-known among these is iterative phase retrieval~\cite{Fienup1982}, most commonly implemented with the well-known Gerchberg-Saxton (GS) algorithm~\cite{Gerchberg1972}. The GS algorithm repeatedly switches between the near- and far-fields by Fourier transform, keeping the near-field phase and neglecting amplitude variations while retaining the far-field phase and replacing its amplitude with the desired CGH pattern. It can be shown that the GS algorithm implements gradient descent on the phase function $\phi(x,y)$, optimizing deviation from a desired far-field intensity pattern in the least-squares sense/, without ever computing a Jacobian~\cite{Fienup1982}.

Unitarity is the matrix analogue of ``phase-only''. Here, we generalize the scalar GS algorithm to operate on matrix quantities using the matrix polar decomposition (for which highly efficient numerical schemes exist) in place of the scalar one. The modified algorithm is shown in Fig. \ref{fig:fig2}. An initial ``near-field'' Jones matrix distribution $\boldsymbol{J}(x,y)$ is chosen and Fourier transformed, yielding $\boldsymbol{A}^{\prime}(k_x, k_y)$. Its polar decomposition is found everywhere, and its Hermitian (lossy) part discarded. From the remaining unitary part, an overall phase $\phi$ is extracted. This is the overall phase of the matrix, and can be chosen as any element of the Jones matrix (such as the upper left element) so long as this choice is consistently applied. The designer-specified, desired far-field polarization behavior, described by $\boldsymbol{A}_{\text{des}}(k_x,k_y)$ is multiplied by this overall phase distribution $\phi(k_x, k_y)$. The resultant quantity is inverse Fourier transformed to yield $\boldsymbol{J}^{\prime}(x,y)$. After a matrix polar decomposition, the unitary part can be extracted, becoming the new ``near-field'' Jones matrix hologram. This cycle continues iteratively until the far field converges to a distribution of Jones matrices that is, ideally, everywhere proportional to $\boldsymbol{A}_{\text{des}}(k_x, k_y)$ up to an overall phase profile $\phi(k_x,k_y)$, a free parameter that evolves upon iteration.

The Fourier transform and the matrix polar decomposition both preserve matrix symmetry. That is, if the desired far-field behavior $\boldsymbol{A}_{\text{des}}(k_x, k_y)=\boldsymbol{A}^T(k_x, k_y)\forall(k_x, k_y)$ (and the initial guess for $\boldsymbol{J}(x,y)$ is as well), no asymmetric matrices are introduced into the iterative scheme of Fig. \ref{fig:fig2}. The resultant $\boldsymbol{J}(x,y)$ will, by definition, be of the form of Eq. \ref{eq:metasurface_jones matrix} so that $\phi_{x^{\prime}}$, $\phi_{y^{\prime}}$, and $\theta$ can be extracted at each point and a metasurface may designed and fabricated in a dielectric platform of choice. The metasurfaces of this work are made of TiO$_2$ pillars for operation at $\lambda=532$ nm using a process documented elsewhere~\cite{Devlin2016}.

In short, the matrix GS algorithm of Fig. \ref{fig:fig2} permits Jones matrix holograms whose far-fields exhibit designer-specified polarization behavior (so long as they obey the above symmetry constraint) to be straightforwardly realized with conventional, pillar-based dielectric metasurfaces. It is a higher-dimensional generalization of previous GS-like algorithms used for polarization-dependent metasurface design, enabled by the matrix polar decomposition. 

\section{Experimental Demonstration}
\label{sec:experiment}

The preceding discussion has been hypothetical: What desired polarization functionality $\boldsymbol{A}_{\text{des}}(k_x, k_y)$ should a metasurface be designed to implement as a proof-of-concept? By the polar decomposition (Eq. \ref{eqn:matrix_polar}), any Jones matrix can be decomposed into a Hermitian (polarizer-like, amplitude-modulating) and a unitary (waveplate-like, phase-modulating) component. This provides a natural categorization of possible experimental test-cases. In what follows, we demonstrate Jones matrix holograms whose far-fields implement both polarizer- and waveplate-like behaviors, beginning with the former. The examples that follow are specifically enabled by the Jones matrix approach of this work (see Sec. \ref{sec:discussion} and supplement S2).

\subsection{Polarization-analyzing holograms}
\label{subsec:pol_analyzing_holograms}

\begin{figure}[!t]
    \centering
    \includegraphics[width=\columnwidth]{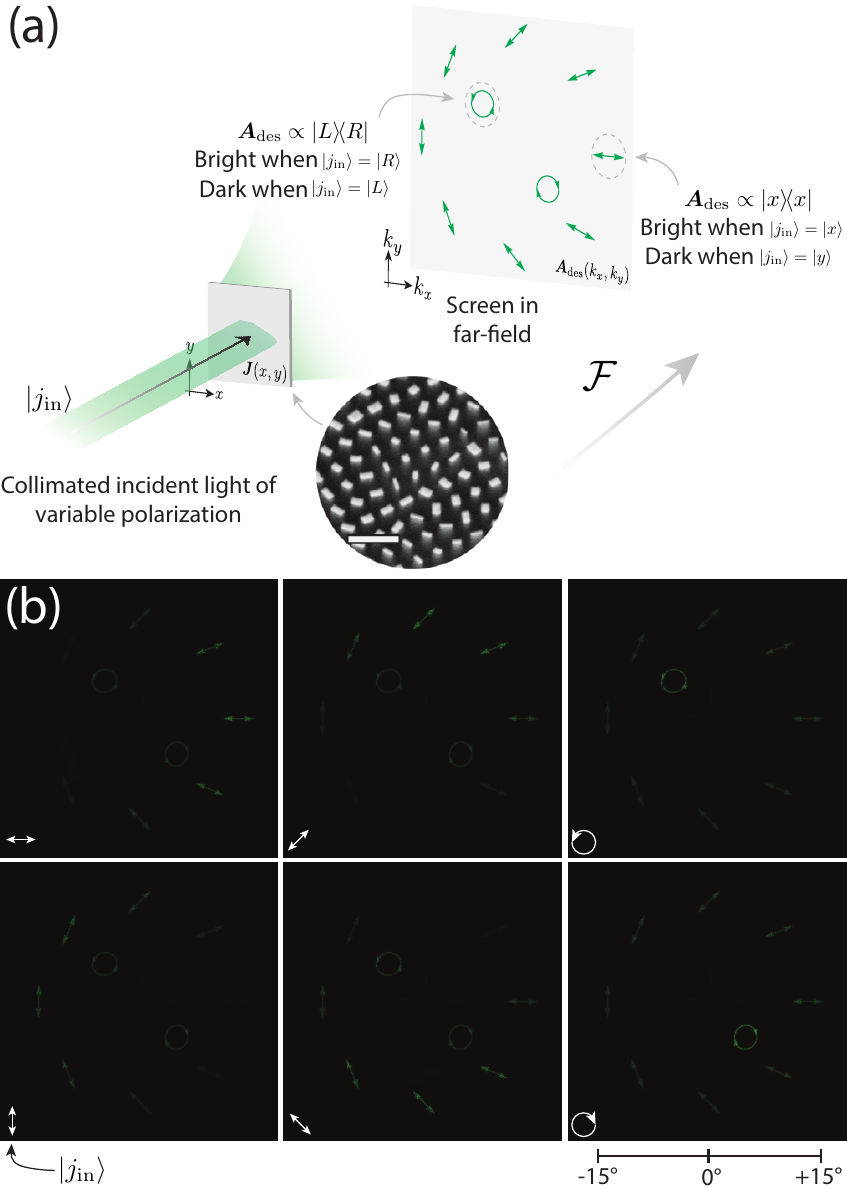}
    \caption{\textbf{Illustrative example of a polarization-analyzing hologram: (a)} When illuminated with collimated laser light, a suitably-designed metasurface hologram implements a far-field in which light is directed on the basis of its incident polarization state. In this particular example, the hologram is designed to produce a pattern of illustrations of different polarization states. Each drawing acts as an ``analyzer'' for its depicted polarization state. For instance, the drawing of $x$-polarized light (horizontal line) is bright when $\ket{j}_{\text{in}}=\ket{x}$ and dark when $\ket{j}_{\text{in}}=\ket{y}$ -- this part of the far-field implements the Jones matrix $\boldsymbol{A}_{\text{des}}=\ketbra{x}{x}$. (SEM scale bar is 1 \textmu m). \textbf{(b)} The far-field measured on a CMOS sensor reflects this desired behavior for six incident polarization states ($\ket{j}_{\text{in}}$ is denoted in white at the lower left of each image). A scale bar shows the cone angle subtended by the far-field.}
    \label{fig:fig3}
\end{figure}

An ideal polarizer passes its preferred polarization state $\ket{\lambda}$ (elliptical in general) without attenuation while extinguishing $\ket{\lambda^{\perp}}$ with $\braket{\lambda}{\lambda^{\perp}}=0$. If this extinction is imperfect, the device is known as a diattenuator~\cite{Chipman2019}. Diattenuators are described by Hermitian Jones matrices.

We seek here to demonstrate Jones matrix holograms whose far-fields implement designer polarizer-like behavior. A conventional polarizer transmits light at its output whose polarization state matches that being analyzed. In other words, a conventional polarizer is of the form

\begin{equation}
\label{eqn:conventional_pol_jm}
    \boldsymbol{A} \propto \ketbra{\lambda}.
\end{equation}

It can be shown that the Jones matrix of Eq. \ref{eqn:conventional_pol_jm} is only symmetric if $\ket{\lambda}$ is a linearly polarization state. The presence of any chirality in the pass polarization $\ket{\lambda}$ destroys the symmetry of the polarizer's Jones matrix and, consequently, its ability to be implemented in the far-field of a conventional pillar-based metasurface Jones matrix hologram (a rule elaborated above). However, it can also be shown that a Jones matrix $\boldsymbol{A} \propto \ketbra{\lambda^*}{\lambda}$, where $*$ denotes complex conjugation, will always be symmetric, irrespective of whether $\ket{\lambda}$ is linear, circular, or elliptical. Such a Jones matrix matches a polarizer's output intensity transfer characteristic, but differs in that the polarization of exiting light is always of flipped-handedness relative to that being analyzed. We refer to this as a polarization-\emph{analyzer}, to distinguish from a true \emph{polarizer}.

In this section, then, we demonstrate devices whose far-fields can be described by the target Jones matrix function

\begin{equation}
\label{eq:polarizer_ff}
    \boldsymbol{A}_{\text{des}}(k_x, k_y) = a(k_x, k_y) \ketbra{\lambda^*(k_x, k_y)}{\lambda(k_x, k_y)}.
\end{equation}

Eq. \ref{eq:polarizer_ff} describes a far-field where each point receives light as though a virtual polarizer were placed there -- the overall amplitude of light directed there $a$ and the incident polarization $\ket{\lambda}$ which evokes maximum intensity can be controlled and may vary arbitrarily from point-to-point.

Fig. \ref{fig:fig3} makes this capability tangible with a simple example. Using the methods above (i.e., matrix phase retrieval), a dielectric metasurface can be designed and fabricated based on a given design. In this case, the metasurface implements a Jones matrix mask $\boldsymbol{J}(x,y)$ whose far-field contains holographic images of different polarization ellipses (eight linear states of varying orientation and both circular polarization states). The region containing each image acts as a polarization analyzer for its respective, depicted polarization state. For example, the holographic image of $\ket{x}$ (horizontal arrow) is brightest when the incident polarization $\ket{j}_{\text{in}}=\ket{x}$ and dark when $\ket{j}_{\text{in}}=\ket{y}$, as though the pixels contained within the drawing act as an analyzer of $\ket{x}$ polarized light. The schematic of the far-field shown in Fig. \ref{fig:fig3}(a) shows all polarization ellipses equally bright for clarity's sake; this would be the case in reality only if the incident light were perfectly unpolarized, with equal projection onto all polarization states.

\begin{figure*}[!ht]
    \centering
    \includegraphics[width=\textwidth]{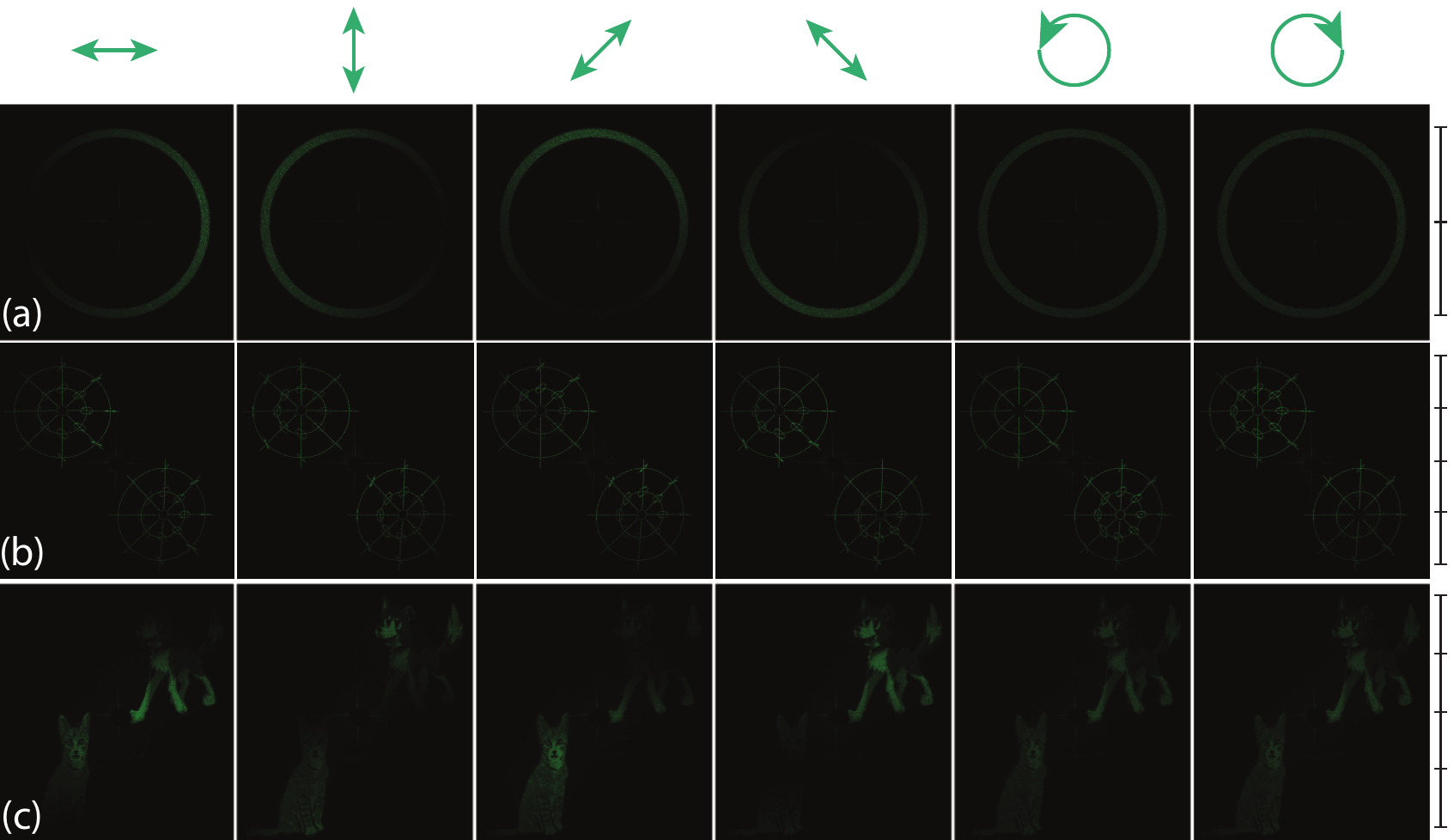}
    \caption{\textbf{Parallel polarization analysis by a metasurface Jones matrix hologram:} Four examples are shown of Jones matrix holograms in which incident light is directed to the far-field in accordance with its projection onto arbitrarily selected analyzer polarizations across the far-field. Each column corresponds to an incident polarization (depicted at the top). Scale bars at the right side of each row show the angular bandwidth of the hologram, with each division corresponding to $15\degree$ centered about $0\degree$. These holograms are measured with a polarimetric relay imaging system described in the supplement (sec. S1D). Animations of the response of hologram (b) to changing input polarization are provided in visualizations \href{http://dx.doi.org/10.6084/m9.figshare.13499298}{1}, \href{http://dx.doi.org/10.6084/m9.figshare.13499304}{2}, and \href{http://dx.doi.org/10.6084/m9.figshare.13499307}{3} (where each number is a clickable link to the public FigShare repository where these animations are hosted; each animation is described at the end of this manuscript).}
    \label{fig:fig4}
\end{figure*}

As shown in Fig. \ref{fig:fig3}(a), the fabricated metasurface is illuminated with collimated laser light ($\lambda=532$ nm in this case) of variable polarization. The angular spectrum (far-field) that results is imaged onto a CMOS sensor using a relay setup described in the supplement (sec. S1D), filtering out the undiffracted zero order along the way. Images are acquired for many input polarization states (without saturation), permitting a full polarimetric characterization of the grating's response.

Fig. \ref{fig:fig3}(b) depicts the far-field produced by the metasurface hologram for six incident polarization states, each of which is denoted in the bottom left corner of its image by a white label. As can be seen, each incident polarization state prompts the strongest response in the region of the hologram corresponding to itself. For example, $\ket{j}_{\text{in}} = \ket{45\degree}$ prompts the hologram to direct most power to the drawing of diagonal, linearly polarized light, while the image of anti-diagonal polarization is dark with a gradient in-between, while drawings of $\ket{x}$, $\ket{y}$, $\ket{R}$, and $\ket{L}$ are all about half as bright. When each circular polarization is incident, all linear polarizations are about equally bright (and half as bright as the image of the incident circular state). The intensity of each polarization depiction is proportional to the projection of the incident polarization state onto the depicted state in accordance with Malus’ Law.  In a way, then,  the hologram of Fig. \ref{fig:fig3} is a “visual full-Stokes polarimeter ” from which an incident polarization state can be simply read out by inspection (this behavior would generalize to partial polarization states as well).

\begin{figure}[!htbp]
    \centering
    \includegraphics[width=\columnwidth]{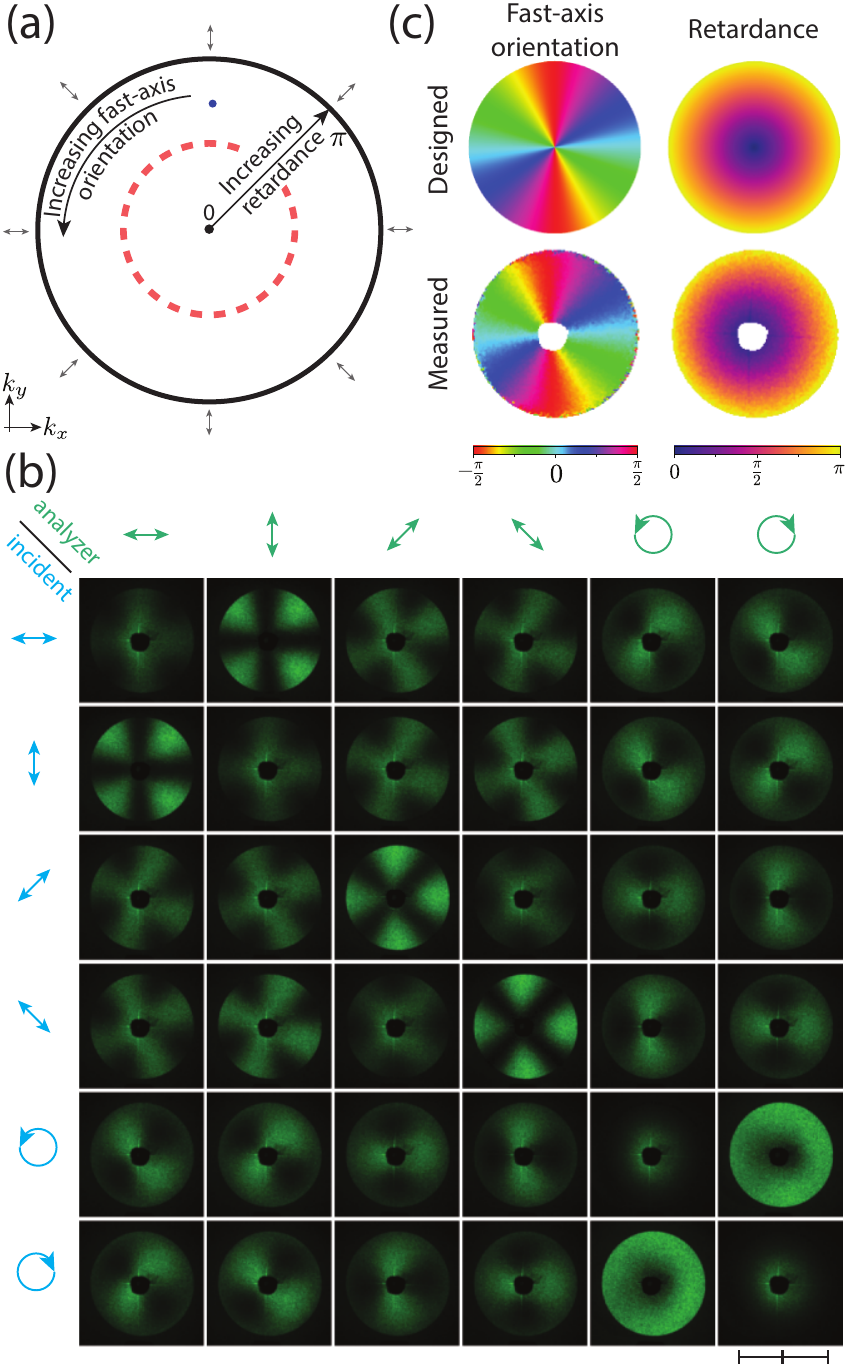}
    \caption{\textbf{Hologram with a far-field exhibiting waveplate-like behavior: (a)} A metasurface is designed whose far-field diffracts light into a circular disk. Each point in this disk is designed to implement a Jones matrix $\boldsymbol{A}_{\text{des}}$ that acts as a birefringent waveplate. The retardance $\Delta$ of this virtual waveplate increases from $0$ to $\pi$ along the radial coordinate while its fast-axis orientation matches the azimuthal coordinate (as denoted by arrows outside the circle). Within this disk all possible linearly birefringent waveplates can be found. For example, the perimeter of the disk represents all possible half-waveplate operations, while the red dotted circle in the center represents the set of all possible quarter-waveplates. A blue dot denotes a 3$\lambda$/8 plate oriented at $90\degree$ as a specific example. \textbf{(b)} The metasurface is illuminated with light of variable input polarization (rows) and its far-field viewed through several polarization analyzers (columns), producing a diverse array of patterns. An angular scale bar denotes the cone angle subtended by the far-field with $15\degree$ divisions. The center of each image is black because the zero-order has been filtered away. \textbf{(c)} Illuminating the metasurface with different input polarization states and imaging its farfield through different polarization analyzers permits reconstruction of the retardance and fast-axis orientation at each point. These closely match design.}
    \label{fig:fig5}
\end{figure}

In Fig. \ref{fig:fig4}, three additional examples showcasing this design freedom are shown. Fig. \ref{fig:fig4}(b) depicts the response of a polarization-analyzing hologram that is, in some sense, a continuous version of Fig. \ref{fig:fig3}. Light is directed into an annular ring with $\boldsymbol{A}(k_x, k_y)$ acting as a linear polarizer that ``turns'' with azimuthal angle. The brightest and darkest parts of the ring rotate with incident linear polarization, always remaining 180$\degree$ apart with Malus' law governing the intensity in-between; for incident circular polarization, the ring appears equally bright everywhere, with half the power of the maxima observed under linearly-polarized illumination.

Fig. \ref{fig:fig4}(c), on the other hand, is a more sophisticated version of the hologram of Fig. \ref{fig:fig3}, depicting a polar projection of the northern and southern hemispheres of the Poincar\'e sphere. Each drawing of a polarization ellipse again acts as an analyzer for its own state. Unlike the example of Fig. \ref{fig:fig4}(a), Fig. \ref{fig:fig4}(c) shows that this approach extends to arbitrary, in-general elliptical polarization states and, moreover, permits features (in particular, the grid lines) that are not sensitive to polarization at all (that is, $\boldsymbol{A}_{\text{des}}\propto\mathbb{I}$ there) to be mixed with those that are. Supplemental Visualizations 1-3 show animations of the response of this hologram to variable linear polarization states, light from a polarizer/rotating quarter-waveplate combination, and partially polarized light, respectively.

Finally, the example of Fig. \ref{fig:fig4}(d) reprises the holographic images from~\cite{Mueller2017} showing that this Jones matrix control can be extended to CGHs with rich features. Here, different parts of the image act as polarization analyzers for linear polarization states. The azimuth of these linear polarization states changes smoothly in the vertical direction from $\ket{y}$ at the bottom, to $\ket{45\degree}$, $\ket{x}$ in the middle, to $\ket{135\degree}$ and finally $\ket{y}$ again at the top. Consequently, the image is uniformly bright (subject to the underlying holographic image) under circularly-polarized illumination.

In each example in Fig. \ref{fig:fig4} (and Fig. \ref{fig:fig3}), a scale bar denotes the extent of the image in angle space with each division denoting $15\degree$ of cone angle (a length scale is not appropriate, as the hologram expands and contracts depending on the screen's placement). The maximum angular bandwidth over which holographic control can be exerted is dictated by the inter-element separation of the metasurface (see supplement sec. S1).

An infinity of other similar examples are possible (including with non-unity diattenuation), subject to energy conservation through a modified form of Parseval's Identity (supplement sec. S1).

\subsection{Waveplate-like holograms}

The examples of the previous section involve a redirection of light depending on its polarization state. A given location in the farfield of these polarization-analyzing holograms receives or does not receive incident light depending on its polarization state, akin to the behavior of a polarizer.

This differs fundamentally from waveplate-like behavior in which output polarization state, rather than intensity, varies with changing input polarization. Waveplates---phase retarders---in contrast, are represented by unitary operators which represent a dephasing of the components of the incident polarization projected onto an orthogonal basis of polarization states (with a possible overall phase shift).

In this section, we demonstrate a Jones matrix hologram whose far-field implements operators of the form

\begin{equation}
\label{eq:retarder_hologram}
    \boldsymbol{A}_{\text{des}}(k_x, k_y) = e^{i\frac{\Delta}{2}}\ketbra{\lambda}+e^{-i\frac{\Delta}{2}}\ketbra{\lambda^{\perp}}
\end{equation}

where $\Delta$ is a retardance angle while $\ket{\lambda}$ and $\ket{\lambda^{\perp}}$ are the eigen-polarizations of the waveplate representing its ``fast-axis''. Eq. \ref{eq:retarder_hologram} projects incoming light into the eigen-basis and retards the two components by the angle $\Delta$ before reconstructing the output. The retardance and eigen-polarization states can vary across the far-field (along with an overall, polarization-independent amplitude). Eq. \ref{eq:retarder_hologram} will always represent a symmetric Jones matrix, and can be implemented in the far-field of a dielectric metasurface, if the eigen-polarizations $\ket{\lambda}$ and $\ket{\lambda^{\perp}}$ represent strictly linear polarization states, as is the case with a conventional birefringent waveplate (as opposed to a crystal possessing, e.g., optical activity).

This capability is demonstrated in Fig. \ref{fig:fig5}. A metasurface is designed to diffract light into a disk in the far-field of uniform intensity. As sketched in Fig. \ref{fig:fig5}(a), each point in the disk is designed to implement a different waveplate operation (cf. Eq. \ref{eq:retarder_hologram}) whose retardance $\Delta$ varies from $0$ to $\pi$ along the radial direction and whose fast-axis direction $\ket{\lambda}$ corresponds to the azimuthal coordinate. For example, the outer edge of the disk is an isoline of all retarders with $\Delta=\pi$, i.e., the set of all half-wave ($\lambda/2$) plates with all possible orientations. The circle halfway between the origin in the edge, denoted by a red dotted line in Fig. \ref{fig:fig5}(a), is similarly the set of all quarter-wave ($\lambda/4$) plates with $\Delta=\pi/2$. All other linearly birefringent waveplates are present, too. The blue dot in Fig. \ref{fig:fig5}(a), for example, denotes a $3\lambda/8$ plate whose fast-axis is oriented at $90\degree$. In other words, the disk, through a proper parameterization of Eq. \ref{eq:retarder_hologram}, contains all linearly birefringent waveplates, at all possible angular orientations, within its extents.

In contrast to the polarization-analyzing holograms of the previous section, the far-field produced by this waveplate-like hologram does not appear to vary (in an intensity sense) when $\ket{j}_{\text{in}}$ changes. However, the changes are revealed when the disk is viewed through a polarization analyzer which is also allowed to vary. Fig. \ref{fig:fig5}(b) shows the patterns produced for six different incident polarization states (again the cardinal polarization states, $\ket{x}$, $\ket{y}$, $\ket{45\degree}$, $\ket{135\degree}$, $\ket{R}$, and $\ket{L}$)) viewed through six polarization analyzers (a turning linear polarizer paired with a $\lambda/4$ plate, for the circular columns). The diverse array of patterns in Fig. \ref{fig:fig5}(b) make intuitive sense when the incident polarization is ``propagated through'' the desired behavior of the far-field as described in Fig. \ref{fig:fig5}(a). For instance, consider the far-field image produced when $\ket{45\degree}$ is incident on the metasurface viewed through an $\ket{R}$ analyzer (second column from right). When diagonally polarized light passes through a quarter-waveplate oriented at $0\degree$, one circular polarization state is produced; if the quarter-waveplate is oriented at $45\degree$, the opposite handedness results. This explains the successive maxima and minima of the images along the quarter-waveplate circle at half the disk's radius, $90\degree$ apart. Theoretical versions of these images, derived from the design, are also given in the supplement for comparison (sec. S3).

Each image in Fig. \ref{fig:fig5}(b) could be similarly intuited. There is, however, a simpler and more direct way of verifying that the far-field behaves in accordance with design. By illuminating the metasurface with a number of different polarization states and viewing these through a number of different polarization analyzers, the $4\times4$ Mueller matrix describing each point in the far-field can be derived. This Mueller matrix maps the Stokes vector of the polarization state incident on the metasurface to the measured polarization state (Stokes vector) at each location in the far-field. The Lu-Chipman Decomposition~\cite{Lu1996} can be applied to each Mueller matrix so that the measured retardance and the fast-axis orientation exhibited by each point in the far-field can be compared to design. This is shown in Fig. \ref{fig:fig5}(c). As designed, the retardance exhibited by the the ``waveplates'' within the far-field disk increases from 0 to $\pi$ at its edge. The orientation rotates smoothly from $0$ to $\pi/2$ and back again twice around the circle (due to the twofold angle degeneracy of polarization states; an azimuthal orientation of $3\pi/2$ is the same as $\pi/2$). In the images acquired of the disk, the center is missing due to undiffracted zero-order light being filtered out in the experimental setup by a physical block. Further details on this measurement and analysis are provided in the supplement (sec. S1D).

\section{Discussion}
\label{sec:discussion}

Dielectric metasurfaces exhibiting form birefringence and, for that matter, a number of other technologies enabling spatially-varying control of optical polarization, already constitute an active area of research with an established literature and are not in themselves a novel contribution of this work. Here, we have instead focused on a way of viewing these optical elements. A description at the level of the Jones calculus permits the polarization operations enacted by individual elements to be mapped to the polarization transfer function of the far-field.

The idea of merging Fourier optics and the Jones calculus in this way does have some precedent (such as~\cite{Moreno2004}, and in the analysis of polarization aberrations of optical systems~\cite{Chipman2019}). However, to our knowledge, it has not been widely applied as a tool for the design of polarization-sensitive diffractive elements (with one notable exception, in which some of us~\cite{Rubin2019} used similar methods based on a brute-force optimization to design polarization-sensitive diffraction \emph{gratings}, rather than holograms). The methods of this work provide a unified mathematical framework that generalizes past work in this area and suggests new possibilities.

For instance, past works have extensively discussed ``polarization switchable'' metasurfaces, that is, metasurfaces which exhibit separate responses or act as an independent optical element depending on the polarization state of illuminating light. In this way, metasurfaces may enable, e.g., lenses that focus in separate locations for $x$ and $y$ polarized light~\cite{Arbabi2015}, holograms with independent far-fields for incident circular polarization of opposite handedness~\cite{Mueller2017}, and gratings directing light to either the $+1$ or $-1$ order depending on which of two orthogonal polarizations is incident. In these works, the polarization basis to which the metasurface is sensitive is fixed across the far-field. The response to a general incident polarization is governed by its projection onto these two chosen basis states.

This work and in particular the polarization-analyzing holograms of Sec. \ref{sec:experiment} shows that this ``switchability'' need not be limited to just two polarization states. Rather than ascribing two global responses to one orthogonal polarization basis, the polarization basis can itself change over the extent of the far-field, as in the holograms of Figs. \ref{fig:fig3} and \ref{fig:fig4}. The ``polarization-analyzing'' behavior exhibited there, as described through the Jones matrix formalism, is a generalization of the ``polarization swichability'' of previous works, one which would not have been possible with previous design strategies. Rather than being limited to just two discrete polarization states, the far-field can ``switch'' on the basis of $N$. This is described in more detail in the supplement (sec. S2). 

Moreover, the ability to enact customizable unitary waveplate-like transformations in the far-field is a possibility overlooked by these previous works. Several previous works enable control of the polarization state of the far-field for a given incident polarization state, e.g., on a set of diffraction orders~\cite{Rubin2018} or over whole holographic images~\cite{Arbabi2019, Deng2018}. 

This work instead suggests that far-field polarization \emph{transformation}, rather than just polarization state, may be controlled.

\section{Conclusion}

In this work, we have introduced Jones matrix holography, in which a polarization-sensitive mask generates a far-field with customizable polarization response. A treatment based on the Jones calculus enables the design and analysis of these holograms without specification of the incident polarization state. Consequently, this generalizes a wide body of past work in polarization-sensitive diffractive optics and metasurfaces. Moreover, the formalism provides a clear link between recent research in this area and extensive work in (scalar) holography of decades past. We have additionally shown how the traditional GS phase retrieval algorithm may be extended to matrix quantities. 

Several possible future directions could broaden this work. For example, the requirement that strictly unitary and symmetric Jones matrix behavior be realized by the far-field could be relaxed with more advanced nanophotonic structures. Metasurfaces composed of, for instance, two layers of dielectric pillars would strongly break mirror symmetry along the $z$ direction, allowing for chiral behavior and thus eliminating the restriction of Jones matrix symmetry. Moreover, lossy structures incorporating, e.g., metals, can implement Hermitian behavior that, when paired with other structures, would afford the designer more freedom to create fully general $\boldsymbol{J}(x,y)$ with mixed Hermitian and unitary responses, perhaps without the use of iterative phase retrieval.

This work may find application in a variety of areas. A Jones matrix hologram could add custom polarization-dependence to an optical system's point spread function, either to addresses systematic polarization aberrations~\cite{Chipman2019} in precision imaging systems or to enable wholly new functionality. Elements based on spatially-varying liquid crystals are already employed in astrophysical measurements for, e.g., exoplanet detection~\cite{otten2014vector}; the present work, and the expanded control it offers, may suggest new possibilities there. Active photonic platforms provide a second promising application area. Light distribution in the holograms of this work is governed by linearity and Malus' Law. Gain-associated nonlinearities in, e.g., a laser cavity could potentially surmount this. This could enable polarization-controlled beamsteering if a far-field pattern such as the annular ring of Fig. \ref{fig:fig4}(a) is used. Finally, this work bears qualitative similarity to research surrounding structured light and optical orbital angular momentum (e.g., the images of Fig. \ref{fig:fig5}(b)), an area where it may find additional interest.

\section*{Animations}

Three animations of the response of the far-field of the metasurface of Fig. 4(b) (which depicts a plot of the north and south poles of the Poincar\'e sphere) to variable input polarization are shown. These animations are hosted on the data repository FigShare. For each, a link and description are provided here. The animations are generated from the measured data from the device.\\

\textbf{Visualization 1} (\href{http://dx.doi.org/10.6084/m9.figshare.13499298}{\underline{\textit{link here}}}): The intensity of the far-field of the ``Poincaré sphere plot'' hologram of Fig. 4(b) can be visualized as the polarization of light incident on the metasurface is varied. The input polarization ellipse in blue and its position on the Poincaré sphere are shown at right. As can be seen, each far-field image of a polarization state has a polarizer-like intensity response to that same polarization state.\\

\textbf{Visualization 2} (\href{http://dx.doi.org/10.6084/m9.figshare.13499304}{\underline{\textit{link here}}}): In this animation, the experimentally observed far-field of the same metasurface is shown as a quarter-waveplate is rotated in front of a horizontally oriented linear polarizer. As shown on the Poincaré sphere, $x$ polarized light evolves into right circular polarization, then $x$, then left circular polarization, and back again, with elliptical states in-between. This is reflected in the brightness of different parts of the far-field - at the appropriate time, each circular polarization sketch goes dark when its orthogonal counterpart is incident on the metasurface. The gridlines appear equally bright, irrespective of incident polarization.\\

\textbf{Visualization 3} (\href{http://dx.doi.org/10.6084/m9.figshare.13499307}{\underline{\textit{link here}}}): Here, the degree of polarization (DOP) of light incident on the metasurface is varied. At the beginning of the animation, light is fully linearly polarized along $x$. Slowly its degree of polarization decreases (while its nominal polarization state remains $x$), which corresponds to the Stokes vector entering the interior of the Poincaré sphere and, eventually, reaching the origin when DOP = 0. As this transition occurs, all parts of the far-field of the Poincaré sphere hologram, schematically representing the incident polarization projected onto a plurality of polarization states, become uniformly bright, since unpolarized light has an equal projection onto all linear polarization states.\\

\section*{Acknowledgments}

The authors thank S. Kheifets for useful discussion. NAR acknowledges support from the National Science Foundation Graduate Research Fellowship Program (GRFP) under grant no. DGE1144152. This work was performed in part at the Center for Nanoscale Systems (CNS), a member of the National Nanotechnology Coordinated Infrastructure (NNCI), which is supported by the National Science Foundation under NSF award no. 1541959. CNS is part of Harvard University. The authors also acknowledge support from the Air Force Office of Scientific Research under award no. FA9550-19-1-0135.

\bibliography{main}

\end{document}